\documentclass{article}

\usepackage[utf8]{inputenc} 
\usepackage[T1]{fontenc}    
\usepackage[colorlinks,allcolors=blue]{hyperref}       
\usepackage{url}            
\usepackage{booktabs}       
\usepackage{amsfonts}       
\usepackage{nicefrac}       
\usepackage{lipsum}
\usepackage{natbib}
\usepackage{enumerate}
\usepackage{float}
\usepackage{gensymb}
\usepackage{amssymb}
\usepackage{amsmath}
\usepackage{graphicx}
\usepackage{mathrsfs}
\usepackage[margin=1in]{geometry}
\usepackage{xcolor}
\usepackage{comment}
\usepackage{enumitem}
\usepackage{amsthm}
\usepackage{bbm}

\usepackage{caption}
\usepackage{subcaption}

\usepackage{setspace}

\newcommand\indep{\protect\mathpalette{\protect\independenT}{\perp}}
\def\independenT#1#2{\mathrel{\rlap{$#1#2$}\mkern2mu{#1#2}}}

\newcommand{\R}{\ensuremath{\mathbb{R}}}

\newcommand{\E}{\ensuremath{\mathbb{E}}}
\newcommand{\calE}{\ensuremath{\mathcal{E}}}

\newcommand{\calX}{\ensuremath{\mathcal{X}}}

\newcommand{\calH}{\ensuremath{\mathcal{H}}}

\theoremstyle{definition}

\theoremstyle{definition}

\theoremstyle{definition}
\newtheorem{theorem}{Theorem} 

\theoremstyle{definition}

\theoremstyle{definition}
\newtheorem{assumption}{Assumption}

\theoremstyle{definition}

\theoremstyle{definition}
\newtheorem{procedure_step}{Step}

\title{Interpretable Sensitivity Analysis for Balancing Weights\thanks{We would like to thank Kevin Guo and Skip Hirshberg for useful discussion and comments. This research was supported in part by the Hellman Family Fund at UC Berkeley, the Institute of Education Sciences, U.S. Department of Education, through Grant R305D200010, the Office of Naval Research (ONR) through grant N00014-17-1-2176, and the Two Sigma PhD fellowship. The opinions expressed are those of the authors and do not represent views of the Institute, the U.S. Department of Education, nor the Office of Naval Research.}}
\author{Dan Soriano, Eli Ben-Michael, Peter Bickel, Avi Feller, and Samuel D. Pimentel \\[0.5em] UC Berkeley and Carnegie Mellon University}
\date{\today}

\begin{document}
\maketitle

\begin{abstract}
\singlespacing

Assessing sensitivity to unmeasured confounding is an important step in observational studies, which typically estimate effects under the assumption that all confounders are measured. 
In this paper, we develop a sensitivity analysis framework for balancing weights estimators, an increasingly popular approach that solves an optimization problem to obtain weights that directly minimizes covariate imbalance.
In particular, we adapt a sensitivity analysis framework using the percentile bootstrap for a broad class of balancing weights estimators. We prove that the percentile bootstrap procedure can, with only minor modifications, yield valid confidence intervals for causal effects under restrictions on the level of unmeasured confounding. 
We also propose an amplification --- a mapping from a one-dimensional sensitivity analysis to a higher dimensional sensitivity analysis --- to allow for interpretable sensitivity parameters in the balancing weights framework. We illustrate our method through extensive real data examples.
\end{abstract}

\onehalfspacing
\clearpage
\section{Introduction}
\label{sec:intro}

Observational studies can be an important source of evidence about causal effects across the medical and social sciences.  Observational studies may be feasible in cases where randomized trials are not, or at least substantially less onerous to conduct at scale, but they raise challenges for analysis that are not present in randomized studies. As one example, consider evaluating the degree to which diets rich in fish elevate blood mercury relative to diets containing little fish. High levels of mercury in the blood can pose health risks; for instance, infants whose mothers had high mercury levels may be at increased risk for adverse neurodevelopmental events \citep{mahaffey2004blood}. Consumption of fish or shellfish has been identified as a major source of mercury in the blood \citep{bjornberg2003methyl}. These effects could be measured by randomly assigning subjects to high- and low-fish diets over long periods of time and comparing their blood mercury, but such experiments may be difficult to conduct and suffer from problems with compliance. Observational data describing blood mercury levels for subjects who choose to eat large or small amounts of fish are more readily available, but direct comparisons between groups are subject to confounding if the high-fish-diet and low-fish-diet subjects are systematically different in other ways. Similarly, measuring the impact of job training programs on wages using randomized experiments is expensive and difficult, but observational studies suffer from substantial confounding \citep{lalonde1986evaluating}.

In observational studies for both examples just described, some confounding may be apparent in the form of obvious differences in observed variables between comparison groups, and analysis often proceeds 
under a key assumption that 
all confounders are measured, sometimes known as \emph{ignorability} or \emph{unconfoundedness}.  
However, this assumption is not verifiable from observed data, and it is often easy to suggest unmeasured  factors that may contribute at least a limited amount of confounding.  For example, in the case of job training programs, one might wonder if individuals who choose to participate in job training may have higher intrinsic motivation to succeed than those who choose not to.
A sensitivity analysis seeks to determine the magnitude of unobserved confounding required to alter a study's findings. If a large amount of confounding is needed, then the study is robust, enhancing its reliability. Assessing sensitivity to unmeasured confounding is a critical part of the workflow for causal inference in observational studies.

In this paper, we develop a sensitivity analysis framework for \emph{balancing weights estimators}. Building on classical methods from survey calibration, these estimators find weights that minimize covariate imbalance between a weighted average of the observed units and a given distribution, such as by re-weighting control units to have a similar covariate distribution to the treated units. Balancing weights have become increasingly common within causal inference, with better finite sample properties than traditional inverse propensity score weighting (IPW). See Section \ref{sec:weighting_estimators} for additional details and \citet{ben2021balancing} for a recent review.

Our proposed sensitivity analysis framework adapts the percentile bootstrap sensitivity analysis that \citet{zhao2019sensitivity} develop for traditional IPW. Specifically, for a given sensitivity parameter, we compute the upper and lower bounds of our estimator for each bootstrap sample, and then form a confidence interval using percentiles across bootstrap samples. 
We prove that this approach yields valid confidence intervals for our proposed sensitivity analysis procedure over a broad class of balancing weights estimators. 

To make a sensitivity analysis more interpretable, \citet{rosenbaum2009amplification} introduce an \textit{amplification} of a sensitivity analysis, which is a mapping from each point in a low-dimensional sensitivity analysis to a set of points in a higher-dimensional sensitivity analysis that all have the same possible inferences. We propose a new amplification that expresses the bias from confounding in terms of: (1) the imbalance in an unobserved covariate; and (2) the strength of the relationship between the outcome and the unobserved covariate. Researchers can then relate the results of our amplification to estimates from observed covariates. 
We demonstrate this approach via a numerical illustration and via several applications.

\section{Background, notation, and review}
\label{sec:framework}

\subsection{Setup and review of marginal sensitivity model}
We consider an observational study setting with independently and identically distributed data $(Y_i, X_i, Z_i)$, $i \in \left\{ 1,\ldots,n \right\}$, drawn from some joint distribution $P(\cdot)$ with outcome $Y_i\in\mathbb{R}$, covariates $X_i \in \calX$, and treatment assignment $Z_i \in \{0,1\}$. We posit the existence of \textit{potential outcomes}: the outcome had unit $i$ received the treatment, $Y_i(1)$, and the outcome had unit $i$ received the control, $Y_i(0)$ \citep{neyman1923, rubin1974estimating}. 
We assume stable treatment and no interference between units \citep{rubin1980randomization}, so
the observed outcome is $Y_i=(1-Z_i)Y_i(0)+Z_iY_i(1)$. 
An estimand of interest is the \textit{Population Average Treatment Effect} (PATE):
\begin{align} \label{eq:ate}
    \tau = \mathbb{E}[Y(1)-Y(0)] = \mu_1 - \mu_0,
\end{align}
where $\mu_1 = \mathbb{E}[Y(1)]$ and $\mu_0 =  \mathbb{E}[Y(0)]$.
To simplify the exposition, we will focus on estimating $\mu_1$; estimating $\mu_0$ is symmetric. We consider an alternative estimand, the \emph{Population Average Treatment Effect on the Treated} (PATT) in Section \ref{sec:examples} and Appendix \ref{sec:att}.

A common set of identification assumptions in this setting, known as \textit{strong ignorability},
assumes that conditioning on the covariates $X$ sufficiently removes confounding between treatment $Z$ and the potential outcomes $Y(0), Y(1)$, and that treatment assignment is not deterministic given $X$ \citep{rosenbaum1983central}.
\begin{assumption}[Ignorability] 
\label{a:ignore}
$Y(0),Y(1) \indep Z \mid X$.
\end{assumption}
\begin{assumption}[Overlap] 
\label{a:overlap} The \emph{propensity score} $\pi(x) \equiv P(Z = 1 \mid X= x)$ satisfies $0<\pi(x)<1$ for all $x \in \calX$.
\end{assumption}
\noindent Under Assumptions \ref{a:ignore} and \ref{a:overlap}, we can non-parametrically identify $\mu_1$, solely with the outcomes from units receiving treatment,
\begin{align} \label{eq:ipw_identity}
  \mu_1 = \E\left[\frac{ZY}{\pi(X)} \right].
\end{align}

In an observational setting, the researcher does not know the \emph{true} treatment assignment mechanism, $\pi(x, y) \equiv P(Z = 1 \mid X = x, Y(1) = y)$, which in general can depend on \emph{both} the covariates $X$ and the potential outcomes $Y(1)$ and $Y(0)$.
A rich literature assesses the sensitivity of estimates to violations of the ignorability assumption. This approach dates back at least to \citet{cornfield1959smoking}, who conducted a formal sensitivity analysis of the effect of smoking on lung cancer. More recent examples of sensitivity analysis include \citet{rosenbaum1983assessing}, \citet{Rosenbaum2002}, \citet{vanderweele2017sensitivity},  \citet{franks2019flexible}, \citet{tudball2019sample}, \citet{Cinelli2020}, \citet{fogarty2020studentized}, \citet{huang2022sensitivity}, and \citet{huang2022variance}. See \citet{hong2020sensitivity} for a recent discussion of weighting-based sensitivity methods.

We adopt the marginal sensitivity model proposed originally by \citet{tan2006distributional} and further developed by \citet{zhao2019sensitivity} and \citet{dorn2021sharp} for traditional IPW weights. 
Following these authors, we split the problem into two parts: sensitivity for the mean of the treated potential outcomes and sensitivity for the mean of the control potential outcomes; without loss of generality, we consider the mean for the treated potential outcomes. Since unbiased estimation of $\mathbb{E}[Y(1)]$ requires knowledge only of $\pi(x,y) = P(Z=1 \mid X = x, Y(1) = y)$ rather than the full propensity score that also conditions on $Y(0)$, we can rewrite Assumption \ref{a:ignore} as $\pi(x, y) = \pi(x)$. 
For details on combining sensitivity analyses for $\mathbb{E}[Y(1)]$ and $\mathbb{E}[
Y(0)]$ into a single sensitivity analysis for the ATE, see Section 5 from \citet{zhao2019sensitivity}.

The marginal sensitivity model
relaxes the ignorability assumption so that the odds ratio between the two conditional probabilities $\pi(x)$ and $\pi(x,y)$ is bounded. 
\begin{assumption}[Marginal sensitivity model]
  \label{a:marginal_sens_model}
  For $\Lambda \geq 1$, the true propensity score satisfies
  \[
    \pi(x, y) \in \calE(\Lambda) = \left\{\pi(x, y) \in (0,1): \Lambda^{-1} \leq \text{OR}(\pi(x), \pi(x,y)) \leq \Lambda \right\},
  \]
  where $\text{OR} (p_1, p_2) = \frac{p_1/(1-p_1)}{p_2/(1-p_2) }$ is the odds ratio.\footnote{
  \citet{zhao2019sensitivity} introduce an extension to the marginal sensitivity model that they call the parametric marginal sensitivity model. The parametric marginal sensitivity model replaces $\pi(x)$ with the best parametric approximation to $\pi(x)$, $\pi_\beta(x)$, and compares $\pi(x,y)$ to $\pi_\beta(x)$ so that the sensitivity analysis addresses both model misspecification and unobserved confounding.}
\end{assumption}
\noindent Here, $\Lambda$ is a sensitivity parameter, quantifying the difference between the true propensity score $\pi(x, y)$ and the probability of treatment given $X = x$, $\pi(x)$;
when $\Lambda = 1$, the two probabilities are equivalent, and Assumption \ref{a:ignore} holds. 
If, for example, $\Lambda = 2$, Assumption \ref{a:marginal_sens_model} constrains the odds ratio between $\pi(x)$ and $\pi(x,y)$ to be between $\frac{1}{2}$ and 2.

Again following \citet{zhao2019sensitivity}, we will consider an equivalent characterization of the set $\calE(\Lambda)$ in terms of the log odds ratio $h(x,y) = \log \text{OR}(\pi(x), \pi(x,y))$:
\begin{equation}
  \label{eq:log_sens_model}
  \calH(\Lambda) = \left\{h : \calX \times \R \to \R : \|h\|_\infty \leq \log \Lambda \right\},
\end{equation}
where $\|h\|_\infty = \sup_{x\in\calX,y\in\R}|h(x,y)|$ is the supremum norm. Rearranging the definition of $h(x,y)$ to be $\log\frac{\pi\left(x,y\right)}{1-\pi\left(x,y\right)} = \log\frac{\pi\left(x\right)}{1-\pi\left(x\right)} - h\left(x,y\right)$ and applying the inverse logit transformation, we can write the true propensity score under a particular sensitivity model $h$ as
\begin{equation}
    \label{eq:shifted_pscore}
    \pi^{(h)}(x,y) = \left[1 + \left( \frac{1}{\pi(x)}  - 1\right)e^{h(x, y)}\right]^{-1}.
\end{equation}
\citet{zhao2019sensitivity} refer to $\pi^{(h)}(x,y)$ as the \emph{shifted propensity score}. Then, for a particular $h \in \calH(\Lambda)$, we can write the \emph{shifted estimand} as
\begin{equation}
    \label{eq:shifted_estimand}
    \mu_1^{(h)} = \E\left[\frac{Z}{\pi^{(h)}(X,Y(1))} \right]^{-1} \E\left[\frac{ZY}{\pi^{(h)}(X,Y(1))} \right].
\end{equation}
Under the marginal sensitivity model in Assumption \ref{a:marginal_sens_model}, we then have a non-parametric partial identification bound, $\inf_{h \in \calH(\Lambda)} \mu_1^{(h)} \leq \mu_1 \leq \sup_{h \in \calH(\Lambda)} \mu_1^{(h)}$. 

The bound just given depends on population quantities that must be estimated, and in practice it is important to take sampling uncertainty into account. \citet{zhao2019sensitivity} use the percentile bootstrap to build confidence intervals that cover this partial identification set, under the assumption that the weights are constructed using IPW.  

We go beyond  \citet{zhao2019sensitivity}'s work in two important ways.  In Section \ref{sec:sensitivity}, we show that the percentile bootstrap strategy for constructing confidence intervals is valid for the broader class of balancing weights, not just IPW.  This requires a different proof strategy than the one based on Z-estimation used by \citet{zhao2019sensitivity} in order to handle balancing weights estimators that achieve approximate (rather than exact) balance on covariates, such as the stable balancing weights of \citet{zubizarreta2015stable}. 
In Section \ref{sec:interpretation} we then introduce an amplification that allows us to better interpret and calibrate marginal sensitivity analyses.

\subsection{Weighting estimators under strong ignorability}
\label{sec:weighting_estimators}

\label{sec:bal_overview}
We estimate $\mu_1$ via a weighted average of treated units' outcomes using weights $\hat{\gamma}(X)$,
\begin{equation}
  \label{eq:mu1_hat}
  \hat{\mu}_1 = \sum_{i=1}^n \frac{Z_i \hat{\gamma}(X_i)}{\sum\limits_{i=1}^n Z_i \hat{\gamma}(X_i)} Y_i.
\end{equation}

\noindent Under strong ignorability (Assumptions \ref{a:ignore} and \ref{a:overlap}), traditional Inverse Propensity Score Weighting (IPW) first models the propensity score, $\hat{\pi}(x)$, directly and then sets weights to be $\hat{\gamma}(X_i) = \frac{1}{\hat{\pi}(X_i)}$. 
Thus, $\hat{\mu}_1$ is a plug-in version of Equation \eqref{eq:ipw_identity}. 
This approach can perform poorly in moderate to high dimensions or when there is poor overlap and either $\pi(x)$ or $\hat{\pi}(x)$ is near 0 or 1 \citep{kang2007demystifying}.

Balancing weights, by contrast, directly optimize for covariate balance; recent proposals include 
\citet{hainmueller2012entropy, zubizarreta2015stable, athey2018approximate, wang2019minimal, hirshberg2019minimax, Tan2020} and have a long history in survey calibration for non-response \citep{Deville1992, Deville1993}.  
See \citet{Chattopadhyay2019} and \citet{ben2021balancing} for recent reviews.

Most balancing weights estimators attempt to control the imbalance between the weighted treated sample and the full sample in some transformation of the covariates $\phi:\calX \to \R^d$.
For example, \citet{zubizarreta2015stable} proposes \emph{stable balancing weights} (SBW) that find weights $\hat{\gamma}(X)$ that solve
\begin{equation} \label{eq:opt_primal}
\begin{aligned}
\min_{\gamma(X) \in \R^{n_1}}
&\quad \int Z \gamma(X)^2 \, dP_n\\
\text{subject to } & \left\|\int Z \gamma(X) \phi(X) - \phi(X) \, dP_n\right\|_\infty \leq \lambda \;\;\;\; \gamma(X) \geq 0,
\end{aligned}
\end{equation}
where $P_n$ is the empirical distribution corresponding to a sample of size $n$ from joint distribution $P(\cdot)$.
These are the weights of minimum variance that guarantee \emph{approximate balance}: 
that the worst imbalance in $\phi$, the transformed covariates, is less than some hyper-parameter $\lambda$. 
There are many other choices of both the penalty on the weights and the measure of imbalance.\footnote{Other possibilities include soft balance penalties rather than hard constraints \citep[e.g.][]{benmichael2020_lor, Keele2020_hosp} and non-parametric measures of balance \citep[e.g.][]{hirshberg2019minimax}.}
For instance, in low dimensions, setting $\lambda = 0$ guarantees \emph{exact balance} on the covariates $\phi(X_i) $. Here we focus on the more common case in which achieving exact balance is infeasible; in that case, the particular choice of penalty function is less important.

The balancing weights procedure is connected to the modeled IPW approach above through the Lagrangian dual formulation of optimization problem \eqref{eq:opt_primal}. The imbalance in the $d$ transformations of the covariates induces a set of Lagrange multipliers $\beta \in \R^d$, and the Lagrangian dual is
\begin{equation}
  \label{eq:opt_dual}
  \min_{\beta \in \R^d} \underbrace{\int Z \left[\beta \cdot \phi(X)\right]_+^2 - \beta \cdot \phi(X) \, dP_n}_{\text{balancing loss}} + \underbrace{\lambda \|\beta\|_1}_{\text{regularization}},
\end{equation}
where $[x]_+ = \max\{0, x\}$. The weights are recovered from the dual solution as $\hat{\gamma}(X_i) = \left[\hat{\beta} \cdot \phi(X_i) \right]_+$.
As \citet{Zhao2019} and \citet{wang2019minimal} show, this is a regularized $M$-estimator of the propensity score when it is of the form $\frac{1}{\pi(x)} = \left[\beta^\ast \cdot \phi(x)\right]_+$ for some true $\beta^\ast$.
Therefore, we can view $\beta^\ast \cdot \phi(x)$ as a natural parameter for the propensity score; different penalty functions will induce different link functions, see \citet{wang2019minimal}.
Similarly, different measures of balance will induce different forms of \emph{regularization} on the propensity score parameters.
In the succeeding sections, we will use this dual connection to show that the percentile bootstrap sensitivity procedure proposed by \citet{zhao2019sensitivity} for traditional IPW estimators in the marginal sensitivity model is valid with balancing weights estimators.

\section{Sensitivity analysis for balancing weights estimators}
\label{sec:sensitivity}

We now outline our procedure for extending the percentile bootstrap sensitivity analysis to balancing weights. We introduce the shifted balancing weights estimator, detail the bootstrap sampling procedure, and describe how to efficiently compute the confidence intervals. Key to constructing the confidence intervals for the partial identification set will be to construct intervals for each sensitivity model $h$ in the collection of sensitivity models $\mathcal{H}(\Lambda)$ in Equation \eqref{eq:log_sens_model}. Each $h$ represents a particular deviation from ignorability that remains in the set defined by the marginal sensitivity model.  We show that the percentile bootstrap yields valid confidence intervals for each sensitivity model in $\mathcal{H}(\Lambda)$, resulting in a valid interval for the partial identification set.   While the procedure for constructing confidence intervals given the weights computed in each bootstrap sample is the same as that in \citet{zhao2019sensitivity},  our result allows for the weights to be constructed by more general methods. 
We provide guidance for interpreting our sensitivity analysis procedure in Section \ref{sec:interpretation}.

To construct the confidence intervals, we first consider the case where we know the log odds function $h(x, y) \in \calH(\Lambda)$. With $h$, we can shift the balancing weights estimator for the shifted estimand $\mu_1^{(h)}$ as
 \begin{align} \label{eq:shifted_est}
     \hat{\mu}_1^{(h)} = \left(\sum\limits_{Z_i=1} \hat{\gamma}^{(h)}(X_i, Y_i(1)) \right)^{-1} \sum\limits_{Z_i=1} \hat{\gamma}^{(h)}(X_i, Y_i(1)) Y_i,
 \end{align}
 where $\hat{\gamma}^{(h)}(X_i, Y_i(1)) = 1 + ( \hat{\gamma}(X_i)  - 1)e^{h(X_i, Y_i(1))}$ for $i \in \{i:Z_i=1\}$ are the shifted balancing weights. Note that there is no requirement for the shifted balancing weights to balance the transformed covariates $\phi$.
We then take $B$ bootstrap samples of size $n$ without conditioning on treatment assignment --- so the number of units in the treatment and control groups may vary from sample to sample --- and re-estimate the weights in each sample by solving the balancing weights optimization problem (\ref{eq:opt_primal}) using the bootstrapped data.

Then, for every $h \in H(\Lambda)$, we can construct a confidence interval for $\mu_1^{(h)}$ using the percentile bootstrap as
\begin{align} \label{eq:conf_int_h}
\left[L^{(h)}, U^{(h)}\right] = \left[Q_{\frac{\alpha}{2}}\left(\hat{\mu}_{1,b}^{*(h)}\right), Q_{1-\frac{\alpha}{2}}\left(\hat{\mu}_{1,b}^{*(h)}\right)\right].
\end{align}
$Q_\alpha(\hat{\mu}_{1,b}^{*(h)})$ is the $\alpha$-percentile of $\hat{\mu}_{1,b}^{*(h)}$ in the bootstrap distribution made up of the $B$ bootstrap samples and $\hat{\mu}_{1,b}^{*(h)}$ is the shifted balancing weights estimator (\ref{eq:shifted_est}) using bootstrap sample $b \in \left\{1, \ldots, B \right\}$. Note, the $^*$ in $\hat{\mu}_{1,b}^{*(h)}$ indicates that it is an estimate from bootstrap data and $b$ is used as an index for the $B$ bootstrap samples.   The following theorem states that $[L^{(h)},U^{(h)}]$ is an asymptotically valid confidence interval for $\mu_1^{(h)}$ with at least $(1-\alpha)$-coverage under high-level assumptions in Appendix \ref{sec:proof_thm1} on how well the balancing weights estimate the propensity scores.

\begin{theorem} \label{theorem:bal_weights}
Under Assumption \ref{assumption:gamma_star} in Appendix \ref{sec:proof_thm1}, for every $h\in H(\Lambda),$
\begin{align*}
    \underset{n\to\infty}{\lim\sup}\;\mathbb{P}_0(\mu_1^{(h)}<L^{(h)})\leq\frac{\alpha}{2}
\end{align*}
and
\begin{align*}
    \underset{n\to\infty}{\lim\sup}\;\mathbb{P}_0(\mu_1^{(h)}>U^{(h)})\leq\frac{\alpha}{2},
\end{align*}
where $\mathbb{P}_0$ denotes the probability under the joint distribution of the data $P(\cdot)$. The probability statements apply under both the conditions on the inverse probabilities and the outcomes in Assumption \ref{assumption:gamma_star} and the marginal sensitivity model (\ref{a:marginal_sens_model}).
\end{theorem}

Since each of the confidence intervals $[L^{(h)}, U^{(h)}]$ are valid, we can use the Union Method to combine them into a single valid confidence interval $[L^{\text{union}}, U^{\text{union}}]$ for $\mu_1$ under Assumption \ref{a:marginal_sens_model}, where
\begin{align} \label{eq:conf_int_union}
L^{\text{union}} = \underset{h\in\mathcal{H}(\Lambda)}{\inf} L^{(h)}, \quad U^{\text{union}} = \underset{h\in\mathcal{H}(\Lambda)}{\sup} U^{(h)}.
\end{align}
Finding $[L^{\text{union}}, U^{\text{union}}]$ would require conducting a grid search over the space of log-odds functions $\calH(\Lambda)$ and computing percentile bootstrap confidence intervals at each point; this is computationally infeasible.
Instead, we can obtain a confidence interval $[L, U]$ for $\mu_1$ by using generalized minimax and maximin inequalities as
\begin{align} \label{eq:zhao_ci}
\left[L, U\right] = \left[Q_{\frac{\alpha}{2}}\left(\underset{h\in\mathcal{H}(\Lambda)}{\inf}\hat{\mu}_{1,b}^{*(h)}\right), Q_{1-\frac{\alpha}{2}}\left(\underset{h\in\mathcal{H}(\Lambda)}{\sup}\hat{\mu}_{1,b}^{*(h)}\right)\right].
\end{align}
\cite{zhao2019sensitivity} show that this interval will be conservative, in the sense of being too wide, since $L \leq L^{\text{union}}$ and $U \geq U^{\text{union}}$.
In fact, \citet{dorn2021sharp} show this can be overly conservative; see Sections \ref{sec:fish} and \ref{sec:discussion} for further discussion.

The extrema of the point estimates can be solved efficiently using Proposition 2 from \citet{zhao2019sensitivity} by the following linear fractional programming problem:
\begin{equation}
\label{eq:weights_lpp}
\begin{aligned}
\underset{r \in \mathbb{R}^{n_1}}{\min/\max}
&\quad  \hat{\mu}_1^{(h)} = \frac{\sum\limits_{i=1}^nZ_i \left(1+r_i \left[\hat{\gamma}(X_i)-1\right]\right)Y_i}{\sum\limits_{i=1}^nZ_i\left(1+r_i \left[\hat{\gamma}(X_i) -1\right]\right)} \\
\text{subject to}
&\quad r_i \in [\Lambda^{-1}, \Lambda],\text{ for all }i\in \left\{1,\ldots,n\right\},
\end{aligned}
\end{equation}
where $r_i= \text{OR}\{\pi(X_i), \pi(X_i,Y_i(1))\}$ are the decision variables.
The procedure to obtain confidence interval $[L, U]$ is then:
\begin{procedure_step}
\label{step:bootstrap}
Obtain $B$ bootstrap samples of the data  of size $n$ without conditioning on treatment assignment.
\end{procedure_step}

\begin{procedure_step} 
\label{step:extrema}
For each bootstrap sample $b=1,\ldots,B$, re-estimate the weights and compute the extrema  $\underset{h\in\mathcal{H}(\Lambda)}{\inf}\hat{\mu}_{1,b}^{*(h)}$ and $\underset{h\in\mathcal{H}(\Lambda)}{\sup}\hat{\mu}_{1,b}^{*(h)}$ under the collection of sensitivity models $\mathcal{H}(\Lambda)$ by solving \eqref{eq:weights_lpp}.

\end{procedure_step}

\begin{procedure_step}
\label{step:CI}
Obtain valid confidence intervals for sensitivity analysis:
\begin{align}
L = Q_{\frac{\alpha}{2}}\left(\underset{h\in\mathcal{H}(\Lambda)}{\inf}\hat{\mu}_{1,b}^{*(h)}\right), \quad U = Q_{1-\frac{\alpha}{2}}\left(\underset{h\in\mathcal{H}(\Lambda)}{\sup}\hat{\mu}_{1,b}^{*(h)}\right).
\end{align}
\end{procedure_step}
\noindent Replacing $\hat{\gamma}(X_i)$ in Equation \eqref{eq:weights_lpp} with the inverse of propensity scores estimated by a generalized linear model recovers the procedure from \cite{zhao2019sensitivity}. As in \citet{zhao2019sensitivity}, the added computational cost for additional values of $\Lambda$ is minimal since they do not require a researcher to draw additional bootstrap samples nor re-estimate the weights.

Finally, a researcher must compute a sensitivity value for a given study; see \citet{Rosenbaum2002} for extensive discussion.
Suppose the confidence interval for PATE under ignorability ($\Lambda = 1$) does not contain zero, indicating a statistically significant effect.
As $\Lambda$ increases, allowing for stronger violations of ignorability, the confidence interval will widen and eventually cross zero.
Of particular interest then is the minimum value of $\Lambda$ for which the confidence interval contains zero; we denote this value as $\Lambda^\ast$.\footnote{Similar to the robustness value with $q = 1$ from \citet{Cinelli2020}, researchers can also consider the minimum value of $\Lambda$ for which the point estimate interval contains zero. The point estimate interval can be computed by solving \eqref{eq:weights_lpp} using the full observed data for a particular value of $\Lambda$.}
Thus, we can interpret $\Lambda^*$ as a necessary difference in the odds ratio between the probability of treatment with and without conditioning on the treated potential outcome for which we no longer observe a significant treatment effect.
This represents the degree of confounding required to change a study's causal conclusions, with larger values of $\Lambda^*$ representing more robust estimates. 

Sensitivity analysis may also be useful in cases where the confidence interval under $\Lambda = 1$ is very small and includes zero, indicating no large effect in any direction or bioequivalence in the sense discussed by \citet{brown1995optimal}. In this setting, a researcher may obtain a sensitivity value $\Lambda^*$ by defining a minimal effect size $\iota > 0$ of practical interest and repeating the sensitivity analysis for larger and larger values of $\Lambda$ until the confidence interval includes either $-\iota$ or $\iota$, revealing the degree of confounding needed to mask a practically important effect.  For examples of such sensitivity analyses, see \citet{pimentel2015large, pimentel2020optimal}.

\section{Amplifying, interpreting, and calibrating sensitivity parameters}
\label{sec:interpretation}

In this section, we provide guidance for interpreting the main sensitivity parameter $\Lambda^*$ by ``amplifying'' the sensitivity analyses into a constraint on the product of: (1) the level of remaining imbalance in confounders after weighting; and (2) the strength of the relationship between the confounders and the treated potential outcome.

In order for a confounder to bias causal effect estimates, it must be associated with both the treatment and the outcome. An ``amplification'' enhances a sensitivity analysis's interpretability by allowing a researcher to instead interpret the results of the sensitivity analysis in terms of two parameters: one controlling the confounder's relationship with the treatment and the other controlling its relationship with the outcome \citep{rosenbaum2009amplification}.
Under the marginal sensitivity model in Assumption \ref{a:marginal_sens_model}, the parameter $\Lambda$ controls how far the propensity score conditioned on only observed covariates $\pi(x)$ can be from an oracle propensity score that includes the treated potential outcome $\pi(x,y)$. This odds ratio bound can be difficult to reason about in applied analyses. To aid interpretation, we propose an amplification that expresses the results of our procedure in terms of the imbalance in confounders and the strength of the relationship between the confounders and the treated potential outcome.

For our amplification, we will use $U \in \R$ to represent a latent unmeasured confounding variable, 
standardized to have mean zero and variance 1.\footnote{\citet{dorn2021sharp} similarly consider a general unobserved confounder $U$, of which $U = Y(1)$ is a special case.} We then consider a working model for the conditional expectation of the treated potential outcome, decomposing it into a term involving the observed covariates $X$ and a linear term for the unmeasured confounder $U$:
\begin{equation}
    \label{eq:amplify_model}
    \E[Y(1) \mid X = x, U = u] = f(x) + \beta_u \cdot u.
\end{equation}
This model merely serves as a guide to interpretation, rather than being a true relationship that we are assuming in the primary causal analysis, and is in fact general. As one extreme case, we can consider a situation in which $f(x) = E[Y(1)]$ and the unmeasured confounder $U$ is a standardized version of the treated potential outcome itself, $U = \frac{Y(1)- \mathbb{E}\left[Y(1)\right]}{\text{sd} \left(Y(1)\right)}$; in this case $\beta_u$ is simply equal to the standard deviation of $Y(1)$. More generally, if some of the variation in $Y(1)$ can be explained by observed covariates and by pure additive noise uncorrelated with treatment, $\beta_u$ describes the amount of additional systematic variation contributed by unobserved confounders. Specifically, $\beta_u$ is the difference in expected $Y(1)$ associated with a one-standard-deviation difference in $U$ while holding covariates fixed.  If one is concerned about multiple unobserved confounders, one may also view $U$ as the one-dimensional function of these confounders that best explains the variance in $Y(1)$'s conditional expectation under model \eqref{eq:amplify_model}.

With this model in place, we can decompose the difference between the true expected value of treated potential outcomes $\mu_1$ and the IPW estimand --- i.e., the bias --- into (i) the strength of the unmeasured confounder $U$ in predicting $Y(1)$ beyond the observed covariates, $\beta_u$, and (ii) the imbalance in $U$, $\delta_u$:
\[
\E[Y(1)] - \E\left[\frac{ZY}{\pi(X)}\right] = \beta_u \cdot \underbrace{\left(\E\left[U\right] - \E\left[\frac{ZU}{\pi(X)}\right]\right)}_{\delta_u}.
\]
Note that here we have used the property that $\E[f(X)] = \E[Zf(X)/\pi(X)]$ for all functions $f$.

Now, we can use the partial identification of $\mu_1$ under the marginal sensitivity model in Assumption \ref{a:marginal_sens_model} to find upper and lower bounds for this product under the sensitivity value $\Lambda^\ast$,
\[
\inf_{h \in \calH(\Lambda^\ast)} \mu_1^{(h)} - \E\left[\frac{ZY}{\pi(X)}\right] \leq \beta_u \cdot \delta_u \leq \sup_{h \in \calH(\Lambda^\ast)} \mu_1^{(h)} - \E\left[\frac{ZY}{\pi(X)}\right] .
\]
These are population-level bounds for the highest and lowest possible bias $\beta_u \cdot \delta_u$. To construct finite-sample versions of these bounds, we
bound the bias as the maximum of the absolute values of the highest and lowest possible differences in the estimated values,
\begin{equation}
    \label{eq:amplification}
 |\beta_u \cdot \delta_u| \leq \max\left\{\left|\inf_{h \in \calH(\Lambda^\ast)} \hat{\mu}_1^{(h)} - \hat{\mu}_1 \right|, \left|\sup_{h \in \calH(\Lambda^\ast)} \hat{\mu}_1^{(h)} - \hat{\mu}_1\right|\right\}.   
\end{equation}
Recall that $\hat{\mu}_1$ \eqref{eq:mu1_hat} is a weighted average of treated units' outcomes using weights $\hat{\gamma}\left(X\right)$.


The constrained relationship between the $\beta_u$ and $\delta_u$ allows us to reason about potential unobserved confounders.
To understand this relationship, we compute a curve that maps the value of the bias to different combinations of $\delta_u$ and $\beta_u$ for enhanced interpretation. 
For example, $\left(\delta_u, \beta_u\right) = \left(1.5, 2\right)$ and $\left(\delta_u, \beta_u\right) = \left(1, 3\right)$ are both consistent with a bias of 3.
Reading off this curve allows the researchers to see that for an unmeasured confounder with any given strength in predicting the treated potential outcome beyond the observed covariates, there must be \emph{at least} some level of imbalance after weighting to induce bias.
To explain a given amount of unmeasured confounding bias, an unmeasured confounder strongly predictive of potential outcomes (after controlling for observed covariates) need only be mildly imbalanced after weighting. 
Conversely, an unmeasured confounder with weak predictive strength must be highly imbalanced even after the observed covariates are approximately balanced by the estimated weights.
In Section \ref{sec:examples}, we illustrate our sensitivity analysis procedure and how our amplification can produce more interpretable results.

\section{Numerical examples}
\label{sec:examples}

We now illustrate the sensitivity analysis and amplification procedures 
using two real data examples. 
We consider the situation in which a researcher uses balancing weights to estimate the Population Average Treatment Effect on the Treated (PATT) of a treatment on an outcome of interest; see Appendix \ref{sec:att} for an overview of the PATT in our setting. 
Based on domain knowledge, the researcher believes that the set of observed covariates includes most factors associated with the treatment assignment and the outcome, while leaving open the possibility that there remain relevant unobserved covariates.

To start,
we compute $\Lambda^*$, which represents the confounding required to alter a study's causal conclusions. In order to compute $\Lambda^*$, we compute confidence intervals for a grid of values of $\Lambda$, starting with $\Lambda = 1$ and then considering larger values of $\Lambda$.
If the confidence interval corresponding to $\Lambda = 1$ contains zero, then the effect estimate is not significant, even under ignorability. If the confidence interval for $\Lambda = 1$ does not contain zero, increasing the value of $\Lambda$ causes the confidence intervals to widen and eventually cross zero for some value of $\Lambda$. We set $\Lambda^*$ equal to the minimum value of $\Lambda$ for which the confidence interval includes zero. Since the the percentile bootstrap procedure induces randomness, this value of $\Lambda^*$ is computed with Monte Carlo error. 

We fix the bias equal to the maximum absolute value of the upper and lower bounds on the bias in Equation \eqref{eq:amplification}.
This value is the maximum absolute value of bias possible under the balancing weights sensitivity model with $\Lambda = \Lambda^*$ and is therefore a level of bias required to overturn the study's causal conclusion.
We create contour plots with curves that map the particular value of bias to varying values of $\delta_u$ and $\beta_u$, allowing the bias to be alternatively interpreted in terms of two sensitivity analysis parameters. \citet{veitch2020sense} use the term ``Austen plot'' to describe similar plots.
We include standardized observed covariates on the contour plots, which serve as guides for reasoning about potential unobserved covariates. Our proposed calibration process using observed covariates is intended to provide a broad sense of plausible parameter values, rather than an attempt to obtain precise estimates as a part of a formal benchmarking exercise. See Section \ref{sec:discussion} for further discussion.
Blue points correspond to observed covariates with imbalance prior to weighting, while red points represent post-weighting imbalance. In the PATT setting, the imbalance prior to weighting in a standardized covariate $X$ can be computed as $\frac{1}{\sum\limits_{i=1}^n Z_i} \sum\limits_{i=1}^n Z_i X_i - \frac{1}{\sum\limits_{i=1}^n (1-Z_i)} \sum\limits_{i=1}^n (1-Z_i) X_i$, while the post-weighting imbalance is $\frac{1}{\sum\limits_{i=1}^n Z_i} \sum\limits_{i=1}^n Z_i X_i - \sum\limits_{i=1}^n \frac{(1-Z_i) \hat{\gamma}(X_i)}{\sum\limits_{i=1}^n (1-Z_i) \hat{\gamma}(X_i)}X_i$.
We view the post-weighting imbalance corresponding to the red points as a best-case scenario for potential unobserved covariates --- in general, we expect to achieve better balance in terms of the observed covariates that we directly target than unobserved covariates. 
Conversely, the pre-weighting imbalance represented by the blue points may be more in line with our expectations for unobserved covariates.

\subsection{LaLonde job training experiment}
\label{sec:lalonde}

We re-examine data analyzed by \cite{lalonde1986evaluating} from the National Supported Work
Demonstration Program (NSW), a randomized job training program. Specifically, we use the subset of data from \cite{dehejia1999causal} to form a treatment group and observational data from the Current
Population Survey--Social Security Administration file (CPS1) to form a control group. We consider estimating the effect of the job training program on 1978 real earnings. The covariates for each individual include their age, years of education, race, marital status, whether or not they graduated high school, and earnings and employment status in 1974 and 1975. In total, there are 185 treated units and 15,992 control units.

\begin{figure}
\centering
\begin{subfigure}[t]{.48\textwidth}
  \centering
  \includegraphics[width=\linewidth]{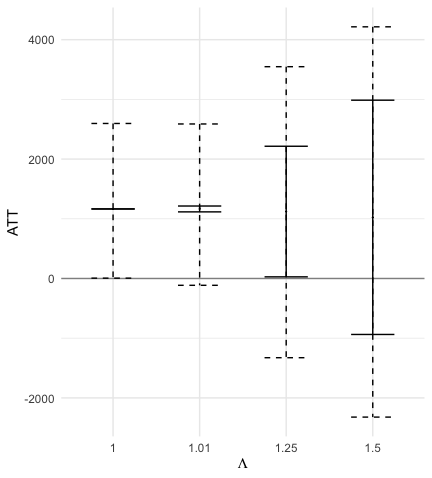}
  \caption{Point estimate and confidence intervals.}
  \label{fig:lalonde_sens_results}
\end{subfigure}%
\hspace{0.5em}
\begin{subfigure}[t]{.48\textwidth}
  \centering
  \includegraphics[width = \linewidth]{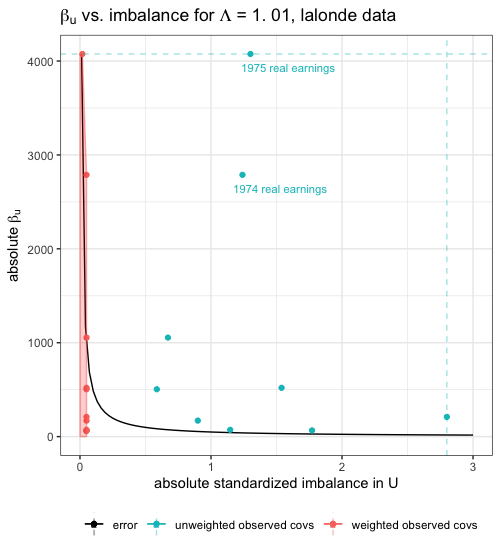}
    \caption{Contour plot illustrating amplification of the sensitivity analysis with comparison to observed variables.}  
    \label{fig:lalonde_contour}
\end{subfigure}
\caption{Sensitivity analysis results with the LaLonde data
\newline (a): Solid intervals are point estimate intervals and dotted intervals are 95\% confidence intervals.
\newline (b): Each location in the plot represents a possible unobserved confounder with parameters $(\delta_u, \beta_u)$ in the amplification.  The contour line gives all such pairs that result in $\Lambda$ equal to the observed sensitivity threshold $\Lambda^* = 1.01$. Plotted points represent observed covariates, with y-coordinates given by absolute multiple regression coefficients in an ordinary least-squares regression of the outcome on standardized covariates among the control group, equivalent to $\beta_u$ if the covariate in question were the only omitted confounder, and with x-coordinates given by treated-control differences in standardized covariates both before weighting (these points are blue) and after weighting (these points are red).  The red shaded region groups locations associated with unobserved confounders no stronger than the observed covariates after weighting, in the sense that some convex combination of post-weighting covariate locations is at least as far from the origin.} 
\label{fig:lalonde_figs}
\end{figure}

First, we use stable balancing weights in Equation \eqref{eq:opt_primal} to estimate
$\widehat{\text{PATT}} = \$1,165$ (estimated with $\phi(x) = x$ and $\lambda = 0.05$), which is in line with \citet{wang2019minimal}'s estimate using slightly different approximate balancing weights.
We then compute $\Lambda^* = 1.01$, which indicates that even a slight difference between the estimated and oracle weights can render the PATT estimate statistically insignificant.
Figure \ref{fig:lalonde_sens_results} shows how the range of point estimates and the 95\% confidence interval widen as $\Lambda$ increases, with the confidence interval including zero for $\Lambda^*$. The range of point estimates is obtained by computing the extrema of the point estimates for a particular $\Lambda$.

Figure \ref{fig:lalonde_contour} shows the contour plot for the LaLonde data, which adds concrete detail to our interpretation of $\Lambda^*$.  The black contour line, representing all combinations of $\beta_u$ and $\delta_u$  for which $\Lambda^* = 1.01$, lies below all of the blue points, suggesting that an unobserved confounder similar even to one of the very weakest observed confounders would be sufficient to reverse the study results.  Furthermore, the black contour line intersects the shaded red region containing post-weighting imbalance, suggesting that even closely-balanced variables like those explicitly accounted for in the weighting algorithm could be sufficient to explain the observed effect. 
All of this strongly substantiates the idea that our study result could be due to very mild unobserved confounding and should not be trusted as a reliable qualitative statement about the true impact of this job training program.  In fact, since several red points lie above the contour line, our finding may even be plausibly explained by residual imbalance in these observed covariates after weighting, whether or not unobserved confounders are present.


Note that visual comparisons of the curve with the blue points and the red region should never be taken at face value as binary statements about whether a study is robust to unmeasured confounding. Instead, one must always account for the context of the individual variables involved.  For instance, the intersection of the curve with the red region occurs only in the upper region of the plot, because two of the variables, real earnings in 1974 and 1975 (both time-lagged versions of the study outcome), are highly correlated with the outcomes.  It is not necessarily plausible that an unobserved confounder would exhibit such high outcome correlation, so intersection with the red region is perhaps less worrying than in a setting where all the observed variables are general demographic measures less directly tied to the observed outcome.  In addition, it is important to include all potentially important observed covariates on the plot lest the red shaded region appear misleadingly small.

\subsection{Fish consumption and blood mercury levels}
\label{sec:fish}

We now examine data analyzed by \cite{zhao2018cross} and \cite{zhao2019sensitivity} from the National Health and Nutrition Examination Survey (NHANES) 2013-2014 containing information about fish consumption and blood mercury levels. We evaluate the sensitivity of estimating the effect of fish consumption on blood mercury levels using balancing weights. There are 234 treated units (consumption of greater than 12 servings of fish or shellfish in the past month) and 873 control units (zero or one servings). The outcome of interest is $\log_2$(total blood mercury), measured in micrograms per liter; the covariates include gender, age, income, whether income is missing and imputed, race/ethnicity, education, smoking history, and the number of cigarettes smoked in the previous month.

\begin{figure}
\centering
\begin{subfigure}[t]{.48\textwidth}
  \centering
  \includegraphics[width=\linewidth]{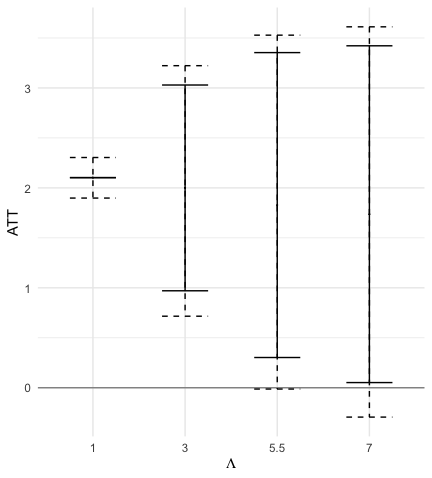}
  \caption{Point estimate and confidence intervals.}
  \label{fig:fish_sens_results}
\end{subfigure}%
\hspace{0.5em}
\begin{subfigure}[t]{.48\textwidth}
  \centering
  \includegraphics[width = \linewidth]{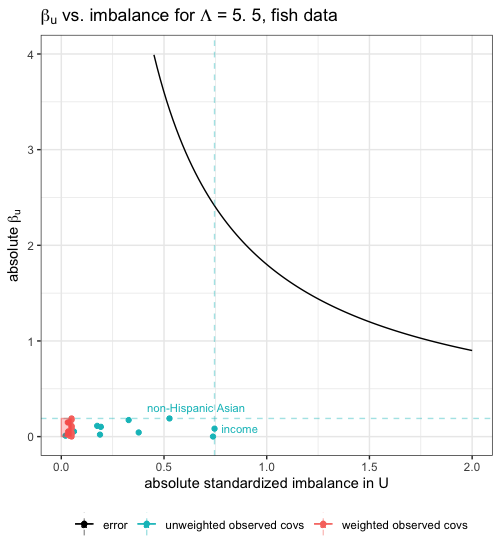}
    \caption{Contour plot illustrating amplification of the sensitivity analysis, with comparison to observed variables.}
    \label{fig:fish_contour}
\end{subfigure}
\caption{Sensitivity analysis results with the fish diet data
\newline (a): Solid intervals are point estimate intervals and dotted intervals are 95\% confidence intervals.
\newline (b): Each location in the plot represents a possible unobserved confounder with parameters $(\delta_u, \beta_u)$ in the amplification.  The contour line gives all such pairs that result in $\Lambda$ equal to the observed sensitivity threshold $\Lambda^* = 5.5$. Plotted points represent observed covariates, with y-coordinates given by absolute multiple regression coefficients in an ordinary least-squares regression of the outcome on standardized covariates among the control group, equivalent to $\beta_u$ if the covariate in question were the only omitted confounder, and with x-coordinates given by treated-control differences in standardized covariates both before weighting (these points are blue) and after weighting (these points are red).  The red shaded region groups locations associated with unobserved confounders no stronger than the observed covariates after weighting, in the sense that some convex combination of post-weighting covariate locations is at least as far from the origin.}
\label{fig:fish_figs}
\end{figure}

To start, the stable balancing weights \eqref{eq:opt_primal} estimate
of the PATT is an increase of 2.1 in $\log_2$(total blood mercury), estimated with $\phi(x) = x$ and $\lambda = 0.05$; $\Lambda^*$ is approximately equal to 5.5 for the fish consumption data. We display the sensitivity analysis results for multiple values of $\Lambda$ in Figure \ref{fig:fish_sens_results}. We observe that the confidence interval corresponding to no confounding ($\Lambda = 1$) is far from zero and that the confidence interval for $\Lambda^* = 5.5$ just begins to cross zero.

The contour plot (Figure \ref{fig:fish_contour}) for the fish data indicates that the causal effect estimate is robust to all but extremely strong unobserved confounders. Here the bias curve is far above the intersection of the dotted lines that represents the maximum strength and pre-weighting imbalance among the observed covariates. Thus, confounding significantly stronger than the observed covariates would be required to alter the causal conclusion.
In particular, consider the most imbalanced pre-treatment confounder, income. The large vertical gap between the associated blue dot (and indeed any of the blue dots) and the contour line suggests that an unobserved confounder sufficient to alter the study's conclusion would not only have to be as imbalanced as income prior to treatment, but would simultaneously have to be a full order of magnitude more predictive of blood mercury than any other variable measured in the study. In fact, in order to change the study's conclusion, an unmeasured confounder as imbalanced as income would have to have an approximately 29 times higher $\beta_u$ than income. While the contour plot itself cannot rule out the possibility that such an unmeasured confounder might exist, it imposes stringent requirements for alternative theories behind the apparent causal effect. 

The LaLonde data results in Figure \ref{fig:lalonde_figs} and the fish consumption data results in Figure \ref{fig:fish_figs} illustrate two extremes for possible outcomes of the sensitivity analysis.  In our experience, more intermediate results frequently arise also; for example, the  contour line might pass above some observed covariates but below others.  In this case especially, it is important to remember that sensitivity analysis is not designed to provide a binary judgment about whether a study's effect is real or not; instead, the contour plot gives a sense for the types of unobserved confounder that might be problematic and the types that can be safely ignored.

\begin{figure}
    \centering
    \includegraphics[width=1\columnwidth]{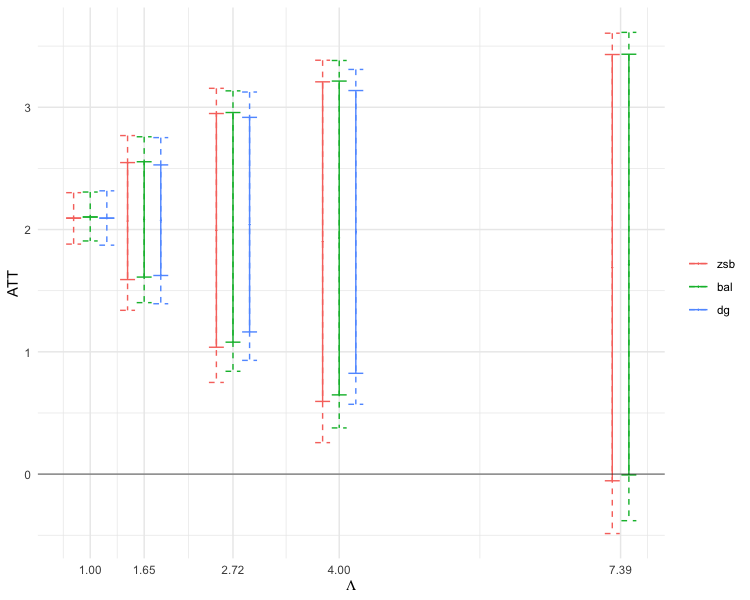}
    \caption{Comparison of confidence interval width after sensitivity analysis for three approaches in the fish consumption example.  We compare the intervals constructed using stabilized weights followed by our proposed sensitivity analysis (labeled ``bal'') against those obtained by fitting IPW weights and conducting sensitivity analysis as described in \citet{zhao2019sensitivity} (``zsb'' in the plot), and against those obtained by IPW and the approach of \citet{dorn2021sharp} (``dg''), at several values of $\Lambda$. The Dorn \& Guo bounds could not be computed at $\Lambda = 7.39$ due to numerical problems encountered in fitting the required quantile regression. All three approaches give similar results, but the balancing weights approach consistently outperforms \citet{zhao2019sensitivity}'s approach, while \citet{dorn2021sharp}'s approach in turn produces narrower intervals than the stabilized weights approach for all values $\Lambda > 1$ investigated.   Note that the results reported here for the \citet{zhao2019sensitivity}
 approach differ slightly from the results reported for their analysis of this dataset because we focus on the ATT rather than the ATE.}
 \label{fig:zhao_dorn_guo_comp}
\end{figure}

Finally, in Figure \ref{fig:zhao_dorn_guo_comp} we compare the results of our sensitivity analysis in the fish consumption data to the results of the approaches described by \citet{zhao2019sensitivity} and \citet{dorn2021sharp}.  As discussed above, \citet{zhao2019sensitivity} use IPW weights and otherwise conduct the sensitivity analysis in an identical manner.   \citet{dorn2021sharp} also use IPW weights but alter the sensitivity analysis by adding a constraint to the population version of the  maximization problem in \eqref{eq:weights_lpp} that enforces balance on certain conditional quantiles of the observed outcomes.  This is designed to ensure that that true propensity scores implied by the sensitivity model balance the observed data properly in large samples (the set of shifted balancing weights over which we take extrema need not do so).
  Figure \ref{fig:zhao_dorn_guo_comp} gives the expanded confidence intervals for the ATT from each approach at three values of $\Lambda$.  All three approaches are qualitatively similar in each case.  However, our approach based on stabilized balancing weights outperforms \citet{zhao2019sensitivity}'s IPW approach at each $\Lambda$-value investigated, achieving strictly shorter intervals.  This suggests that the ability of balancing weights to achieve more precise inference than IPW in moderate samples, previously documented for settings with no unobserved confounding \citep{ben2021balancing}, seems to extend to sensitivity analysis as well.  The approach of \citet{dorn2021sharp} achieves narrower intervals than either of the other approaches; however, we note that \citet{dorn2021sharp}'s added constraint relies on quantile regression and hence requires the outcome to be continuous, unlike the other two approaches. Additionally, the authors find that the quantile balancing confidence intervals can result in under-coverage when the quantiles are correctly specified, which could suggest a setting in which our proposed sensitivity analysis procedure's wider intervals could be advantageous. As such the combination of stabilized balancing weights and sensitivity analysis appears to offer an attractive mix of generality and precision compared to existing competitors.


\section{Discussion}
\label{sec:discussion}

Balancing weights estimation is a popular approach for estimating treatment effects by weighting units to balance covariates. In this paper, we develop a framework for assessing the sensitivity of these estimators to unmeasured confounding. We then propose an amplification for enhanced interpretation and illustrate our method through real data examples.

We briefly outline potential directions for future work. First, as discussed in Section \ref{sec:fish}, \citet{dorn2021sharp} show that the intervals obtained from solving the linear programming problem \eqref{eq:weights_lpp} can be overly conservative, and resolve this issue by adding constraints that require balance on certain conditional quantiles of the outcome. 
It seems likely that such constraints would offer benefits for balancing weights estimators as well.
We leave a thorough investigation to future work.

Second, we could extend our framework to include augmented balancing weights estimators, which use an outcome model to correct for bias due to inexact balance. Additionally, we could extend our sensitivity analysis framework to balancing weights in panel data settings. For example, we could adapt this framework to variants of the synthetic control method \citep{abadie2003economic, ben2018augmented}, extending proposals for sensitivity analysis from \citet{firpo2018synthetic}.

Additionally, \citet{Cinelli2020} point out that informal benchmarking procedures can be misleading if used to perform an exact calibration of sensitivity analysis parameters based on observed data. The authors argue that this occurs because the estimates of the observed covariates' relationships with the outcomes may be impacted by unmeasured confounding. They propose a formal benchmarking procedure to bound the strength of unmeasured confounders based on observed covariates. Adapting \citet{Cinelli2020}'s formal benchmarking procedure to our setting could be a topic of future research.

Finally, we could use our framework to provide guidance in the design stage of balancing weights estimators. When estimating treatment effects using balancing weights, researchers must make decisions including the specific dispersion function of the weights, the particular imbalance measure, and, in many cases, an acceptable level of imbalance. We could extend our sensitivity analysis procedure to help make these decisions to improve robustness and power in the presence of unmeasured confounding. 
For example, we could provide insight into the trade-off between achieving better (marginal) balance on a few covariates or worse balance on a richer set of covariates.


\clearpage
\singlespacing
\bibliographystyle{apalike}
\bibliography{references}

\appendix

\clearpage
\section{Proofs}
\label{sec:app_proofs}

\subsection{Proof of Theorem \ref{theorem:bal_weights}}
\label{sec:proof_thm1}

\begin{proof} We prove that, after centering, the difference between the mean computed from estimating and evaluating the inverse probability function $\gamma$ on bootstrap data and the mean computed from using the true function $\gamma$ and evaluating on actual data is of order $n^{-1/2}$.

For simplicity, we consider estimating the population mean from an independent and identically distributed random sample with missing outcome data. For unit $i$, let $Y_i$ be the outcome, $X_i$ be a vector of observed covariates, and $Z_i$ be a response indicator, where $Z_i = 1$ if we observe unit $i$'s outcome and $Z_i = 0$ otherwise. In addition, let $\gamma_P(X) = 1/\pi_P(X)$ be the \emph{population weight} associated with the unit with covariate $X$.
We consider using estimator $\hat{\mu}^{(h)} = \frac{1}{n}\sum\limits_{i=1}^n\hat{\gamma}^{(h)}(X_i, Y_i)Z_iY_i$ to estimate $\mu^{(h)}=\mathbb{E}[Y]= \mathbb{E}[\mathbb{E}[Y|X, Y]] = \mathbb{E}[\mathbb{E}[\frac{ZY}{\pi_P^{(h)}(X,Y)}|X, Y]] = \mathbb{E}[\frac{ZY}{\pi_P^{(h)}(X, Y)}]=\mathbb{E}[\gamma^{(h)}_P(X, Y)ZY]$ (by the law of iterated expectations) from observed data $O_i = (X_i, Z_i, Y_iZ_i)_{i=1}^n$ drawn from joint distribution $P(\cdot)$. Theorem \ref{theorem:bal_weights} applies for any known deviation from ignorability represented by the log odds ratio $h(x,y) = \log \text{OR}(\pi(x), \pi(x,y))$. Without loss of generality, we use $h(x,y) = 0$ and suppress the dependency of $\hat{\mu}^{(h)}$ and $\mu^{(h)}$ on $h(x,y)$ for notational simplicity.

We sample split to make the proof and arguments simpler and more transparent \citep[see][]{klaassen1987consistent}. The proof can equivalently be done without sample splitting, but we sample split to avoid the associated complexities. We split the data into two equally sized samples, $i = 1,\ldots,m$ and $i = m+1,\ldots,n$. For both samples, we take an iid bootstrap sample of size $m$ from the respective empirical distribution to obtain data $O_i^* = (X_i^*, Z_i^*, Y_i^*Z_i^*)_{i=1}^m$ and $O_i^* = (X_i^*, Z_i^*, Y_i^*Z_i^*)_{i=m+1}^n$. Let $\hat{\gamma}^*$ denote an estimate of $\gamma$ using bootstrap data. We estimate $\hat{\gamma}^*(X)$ in one bootstrap sample and evaluate in the other bootstrap sample. We then switch roles and take a weighted average of the two estimates proportional to $\sum_{i=1}^m Z_i^*$ in both bootstrap samples to obtain an efficient estimate. This sample splitting approach with reversing roles and averaging yields the same estimate as without sample splitting to order $o(n^{-1/2})$. We demonstrate this through simulation (see Appendix \ref{sec:app_sim}).
We examine the case where we evaluate on the bootstrap sample from the second half of the data and estimate $\hat{\gamma}^*(X)$ from the bootstrap sample from the first half.

We make the following mild assumptions on how $\hat{\gamma}$ is constructed:

\begin{assumption} \label{assumption:gamma_star}

Consider function $\Tilde{\gamma}:\mathscr{X}^m \times \{0,1\}^m \to \mathbb{R}^+$. As an example, consider the function corresponding to the stable balancing weights optimization problem \eqref{eq:opt_primal}. Let $\hat{\gamma}_n^*(x) = \Tilde{\gamma}_{P^\ast_m}(X_1^*,\ldots, X_m^*, Z_1^*,\ldots, Z_m^*,  x)$ and $\hat{\gamma}_n(x) = \Tilde{\gamma}_{P_m}(X_1,\ldots, X_m, Z_1,\ldots, Z_m, x)$, where $P^\ast_m$ and $P_m$ are the empirical distributions for the bootstrap sample from the first half of the data and the actual first half of the data, respectively, be such that:
\begin{enumerate}
    \item $\Tilde{\gamma}$ is uniformly bounded in $m$ and $x$.
    \item \label{a:E_sup} $\mathbb{E}_1\left[\left(\underset{x}{\sup}\Big| \hat{\gamma}_n^*(x) - \hat{\gamma}_n(x) \Big|\right)^2 \right] = o_p(1)$.
    \item \label{a:sup} $\underset{x}{\sup}\Big|\hat{\gamma}_n(x) - \gamma_P(x)\Big| = o_p(1)$.\footnote{\cite{wang2019minimal}'s Theorem 2 proves that Assumption \ref{assumption:gamma_star}.\ref{a:sup} holds for weights estimated by SBW \eqref{eq:opt_primal}.}
    \item \label{a:bias} $\mathbb{E}\left[\hat{\gamma}_n(X)ZY \right] - \mathbb{E}\left[\gamma_P(X)ZY \right] = o(n^{-1/2})$.
    \item \label{a:finite_moment} $Y$ has a finite second moment: $\mathbb{E}[Y^2] \leq M$, where $M$ is a constant.
\end{enumerate}
\end{assumption}

Assumption \ref{assumption:gamma_star}.\ref{a:bias} assumes that the bias of $\hat{\mu}$ for estimating $\mu$ is of order $o(n^{-1/2})$. The assumptions for Theorem 3 in \citet{wang2019minimal}  and Theorem 2 in \citet{hirshberg2019minimax} are possible conditions under which our set of assumptions hold. These are representative of typical assumptions in this setting where the estimator is assumed to be a function of $d$ covariates with assumptions on the number of components of an orthogonal expansion. There are various alternative assumptions, all of which boil down to requiring that $\gamma_P(\cdot)$ can be characterized by a low-dimensional structure.

In conjunction with Assumptions \ref{assumption:gamma_star}.\ref{a:E_sup} and \ref{assumption:gamma_star}.\ref{a:sup}, Assumption \ref{assumption:gamma_star}.\ref{a:bias} can be implausible with high-dimensional covariates or when the covariate distribution can be specified only by a high-dimensional parametric model which requires estimation. This caution is independent of the method used to estimate $\gamma_P(X)$. We provide an example to illustrate the issues that can arise in high-dimensional settings. Suppose $X$ has dimension $p$ and that $\hat{\gamma}_n(X)$ uses Nadaraya-Watson type kernel density estimation for $\gamma_P(X)$. Further, assume that $\gamma_P(\cdot)$ has bounded partial derivatives of order $\leq s$. Then, it is well known that if $\hat{\gamma}_n(X)$ has bandwidths $h_1 = \cdots = h_p = h$, then $\mathbb{E}\left[\hat{\gamma}_n(X)|X\right] = \gamma_P(X) + O(h^s)$ and $\mathbb{E}\left[\left|\hat{\gamma}_n(X) - \mathbb{E}\left[\hat{\gamma}_n(X)|X\right] \right|^2\right] = \Omega((nh^p)^{-1})$. In order to have $nh^p \to \infty$ and $nh^{2s} \to 0$, we must have:
\begin{enumerate}
    \item $h \to 0$ slower than $n^{-\frac{1}{p}}$ and
    \item $h \to 0$ faster than $n^{-\frac{1}{2s}}$.
\end{enumerate}
\noindent This is possible only if $s \geq p/2$. In fact, more sophisticated heuristics yield replacement of $\frac{1}{2s}$ by $\frac{1}{4s}$. Intuitively, if $p$ is large, this assumption is unrealistic in any case. It implies that $\gamma_P(\cdot)$ has a Taylor expansion to order $s$ with $\Omega(p^s)$ bounded coefficients, which means $p^{\frac{p}{2}}$ for $s \geq p/2$. For $p = 100$, this yields $100^{50}$! This example illustrates that there is reason to be skeptical of the plausibility of Assumption \ref{assumption:gamma_star}.\ref{a:bias} in high-dimensional settings. Additional research into propensity score estimation with high-dimensional covariates would seem important.

These assumptions together imply that $\hat{\gamma}_n^*$ is consistently uniform for $\gamma$. Assumption \ref{assumption:gamma_star} verifies
\begin{align*}
    & \mathbb{E}_1\Big[ \Big( \underset{x}{\sup}\Big| \hat{\gamma}_n^*(x) - \gamma_P(x) \Big| Y_{m+1} Z_{m+1} \Big)^2 \Big] \\
    =& \mathbb{E}_1\Big[ \Big( \underset{x}{\sup}\Big| \hat{\gamma}_n^*(x) - \gamma_P(x) \Big| \Big)^2\Big] \mathbb{E}_1\Big[Y_{m+1}^2 Z_{m+1}^2 \Big] \\
    =& o(1), 
\end{align*}
where $\mathbb{E}_1$ denotes the conditional expectation given the first sample. Note, the conditions in Assumption \ref{assumption:gamma_star} are stronger than needed and could be relaxed.

We proceed conditional on the first sample $O_i = (X_i, Z_i, Y_iZ_i)_{i=1}^m$ and the first bootstrap sample $O_i^* = (X_i^*, Z_i^*, Y_i^*Z_i^*)_{i=1}^m$. Therefore, $\hat{\gamma}_n^*$ is a completely known function. Let $\mathbb{E}^*$ denote the conditional expectation of the second bootstrap sample given the actual second sample.

Since
\begin{align*}
    \mathbb{E}^*\Big[ \frac{1}{m} \sum\limits_{i=m+1}^{2m} \hat{\gamma}_n^*(X_i^*)Z_i^*Y_i^*\Big] = \frac{1}{m}\sum\limits_{i=m+1}^{2m} \hat{\gamma}_n^*(X_i)Z_iY_i,
\end{align*}
then, by Theorem 2.1 from \citet{bickel1981some},
\begin{align}
    &\frac{1}{m}\sum\limits_{i=m+1}^{2m} \hat{\gamma}_n^*(X_i^*)Z_i^*Y_i^* - \frac{1}{m}\sum\limits_{i=m+1}^{2m} \hat{\gamma}_n^*(X_i)Z_iY_i \label{eq:all_boot_term}\\
    \text{and}\quad& \frac{1}{m}\sum\limits_{i=m+1}^{2m} \Big(\hat{\gamma}_n^*(X_i)Z_iY_i - \mathbb{E}_1\Big[ \hat{\gamma}_n^*(X_{m+1})Z_{m+1}Y_{m+1} \Big]  \Big) \label{eq:boot_fcn_term}
\end{align}
have the same limiting distribution. Since (\ref{eq:all_boot_term}) and (\ref{eq:boot_fcn_term}) have the same limiting distribution, instead of showing
\begin{equation} \label{eq:want_prove_orig}
\begin{aligned}
    &\frac{1}{m}\sum\limits_{i=m+1}^{2m} \hat{\gamma}_n^*(X_i^*)Z_i^*Y_i^* - \mathbb{E}^*\Big[ \frac{1}{m} \sum\limits_{i=m+1}^{2m} \hat{\gamma}_n^*(X_i^*)Z_i^*Y_i^*\Big]  \\
    =& \frac{1}{m}\sum\limits_{i=m+1}^{2m} \gamma_P(X_i)Z_iY_i - \mathbb{E}_1\Big[\gamma_P(X_{m+1})Z_{m+1}Y_{m+1}\Big] + o_p(n^{-1/2})
\end{aligned}
\end{equation}
to show that the bootstrap can be validly applied, it suffices to show that the difference between the mean with the true $\gamma$ and the mean with $\hat{\gamma}_n^*$ estimated on the bootstrap data is of order $n^{-1/2}$. Therefore, we show
\begin{equation} \label{eq:want_prove2}
\begin{aligned}
    &\frac{1}{m}\sum\limits_{i=m+1}^{2m} \hat{\gamma}_n^*(X_i)Z_iY_i - \mathbb{E}_1\Big[\hat{\gamma}_n^*(X_{m+1})Z_{m+1}Y_{m+1}\Big]  \\
    =& \frac{1}{m}\sum\limits_{i=m+1}^{2m} \gamma_P(X_i)Z_iY_i - \mathbb{E}_1\Big[\gamma_P(X_{m+1})Z_{m+1}Y_{m+1}\Big] + o_p(n^{-1/2}).
\end{aligned}
\end{equation}

We have now reduced the problem to showing that the true function $\gamma$ can be replaced with $\hat{\gamma}_n^*$. In order to show this, we use properties of $\hat{\gamma}_n^*$ from Assumption \ref{assumption:gamma_star}. First, we let
\begin{align*}
    \Delta(X_i,Y_i,Z_i) = (\hat{\gamma}_n^*(X_i)-\gamma_P(X_i))Z_iY_i - \mathbb{E}_1 \Big[ (\hat{\gamma}_n^*(X_{m+1})-\gamma_P(X_{m+1}))Z_{m+1}Y_{m+1} \Big].
\end{align*}

Note that the difference between the terms on the left and right hand sides of (\ref{eq:want_prove2}) is equal to $\frac{1}{m}\sum\limits_{i=m+1}^{2m} \Delta(X_{i},Y_{i},Z_{i})$. Additionally, note that $\mathbb{E}_1\Big[\Delta(X_{i},Y_{i},Z_{i})\Big]=0$. Therefore,
\begin{align*}
    &\mathbb{E}_1\Big[ \Big( \frac{1}{m}\sum\limits_{i=m+1}^{2m} \Delta(X_{i},Y_{i},Z_{i})\Big)^2\Big] \\
    =&\frac{1}{m} \mathbb{E}_1 \Big[ \Delta(X_{m+1},Y_{m+1},Z_{m+1})^2 \Big].
\end{align*}
Since $m = \Omega(n)$, by Assumption \ref{assumption:gamma_star},
\begin{align*}
    &\mathbb{E}_1 \left[ \Delta(X_{m+1},Y_{m+1},Z_{m+1})^2 \right] \\
    =&\mathbb{E}_1 \left[ \left( \left[\hat{\gamma}_n^*(X_{m+1})-\gamma_P(X_{m+1})\right]Z_{m+1}Y_{m+1}\right)^2 \right] \\ 
    &- 2\mathbb{E}_1 \Big\{\big[\hat{\gamma}_n^*(X_{m+1})-\gamma_P(X_{m+1})\big]Z_{m+1}Y_{m+1} \mathbb{E}_1 \Big[ \big[\hat{\gamma}_n^*(X_{m+1})-\gamma_P(X_{m+1})\big]Z_{m+1}Y_{m+1} \Big]\Big\} \\ 
    &+\mathbb{E}_1 \Big\{\mathbb{E}_1 \Big[ \big[\hat{\gamma}_n^*(X_{m+1})-\gamma_P(X_{m+1})\big]Z_{m+1}Y_{m+1} \Big]^2\Big\} \\
    =&\mathbb{E}_1 \Big[ \Big( \big[\hat{\gamma}_n^*(X_{m+1})-\gamma_P(X_{m+1})\big]Z_{m+1}Y_{m+1}\Big)^2 \Big] - \mathbb{E}_1 \Big[ \big[\hat{\gamma}_n^*(X_{m+1})-\gamma_P(X_{m+1})\big]Z_{m+1}Y_{m+1} \Big]^2 \\
    \leq& \mathbb{E}_1 \Big[ \Big( \big[\hat{\gamma}_n^*(X_{m+1})-\gamma_P(X_{m+1})\big]Z_{m+1}Y_{m+1}\Big)^2 \Big] \\
    \leq& M \cdot \mathbb{E}_1 \left[ \left(\hat{\gamma}_n^*(X_{m+1})-\gamma_P(X_{m+1})\right)^2\right] \\
=& o_p(1).
\end{align*}
Therefore, (\ref{eq:want_prove2}) follows.
\end{proof}

\clearpage
\section{Simulation for sample splitting}
\label{sec:app_sim}

We conduct simulations to demonstrate the validity of the sample splitting technique that we use to prove Theorem \ref{theorem:bal_weights} in Appendix \ref{sec:proof_thm1}. We show that the bootstrap distributions for the balancing weights estimates of $\mu_0$ with and without sample splitting are quite similar.

\begin{figure}[!htb]
    \centering
    \includegraphics[width = 0.5\linewidth]{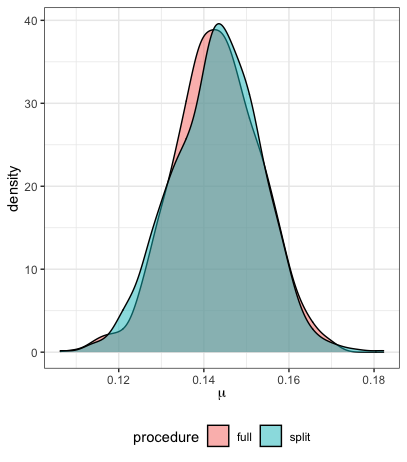}
    \caption{Bootstrap distributions of estimates of $\mu_0$ with the full data and with sample splitting}
    \label{fig:split_sims}
\end{figure}

The setup of the simulations is as follows. We draw 10,000 iid samples where covariates $X_1$ and $X_2$ are drawn from standard normal distributions, treatment indicator $Z_i$ is a bernoulli random variable with probability = $0.5 + 0.07 X_{1i} + 0.07 X_{2i} + \epsilon_i,$ where $\epsilon_i \sim \mathcal{N}(0,0.03^2)$, and $Y_i = 0.2 Z_i + 0.5 X_{1i} + 0.5 X_{2i} + \delta_i,$ where $\delta_i \sim \mathcal{N}(0,0.2^2)$. We run 1,000 simulations and estimate $\mu_0$ with and without sample splitting using weights obtained by entropy balancing with exact balance from \citet{hainmueller2012entropy}. We observe in Figure \ref{fig:split_sims} that the bootstrap distributions of the estimates with and without sample splitting are comparable.

\clearpage
\section{Average treatment effect on the treated}
\label{sec:att}

In many settings, researchers are interested in estimating the \textit{Population Average Treatment Effect on the Treated} (PATT):

\begin{align} \label{eq:att}
    \tau_T = \mathbb{E}[Y(1)-Y(0)|Z=1] = \mu_{11} - \mu_{01},
\end{align}
where $\mu_{11} = \mathbb{E}[Y(1)|Z=1]$ and $\mu_{01} =  \mathbb{E}[Y(0)|Z=1]$. Since $\mu_{11}$ is identifiable from observed data, we primarily focus on estimating $\mu_{01}$.

Our procedure for performing sensitivity analysis outlined in Section \ref{sec:sensitivity} largely still holds. The primary details that differ for the PATT are as follows. First, for a particular $h \in H(\Lambda)$, we can write the shifted estimand as 
\begin{align} \label{eq:shifted_estimand_patt}
    \mu_{01}^{(h)} = \E\left[(1-Z)\frac{\pi^{(h)}(X,Y(0))}{1-\pi^{(h)}(X,Y(0))} \right]^{-1} \E\left[(1-Z)\frac{\pi^{(h)}(X,Y(0))}{1-\pi^{(h)}(X,Y(0))} Y \right].
\end{align}
The corresponding shifted estimator for $\mu_{01}^{(h)}$ is
 \begin{align} \label{eq:shifted_est_patt}
     \hat{\mu}_{01}^{(h)} = \left(\sum\limits_{Z_i=0} e^{-h(X_i, Y_i(0))} \hat{\gamma}(X_i) \right)^{-1} \sum\limits_{Z_i=0} e^{-h(X_i, Y_i(0))} \hat{\gamma}(X_i)  Y_i.
 \end{align}



We make the following modifications to our amplification described in Section \ref{sec:interpretation} for the ATT. Where $U \in \R$ represents a latent unmeasured confounding variable, 
standardized to have mean zero and variance 1, we consider a working model for the conditional expectation of the control potential outcome:
\begin{equation}
    \label{eq:amplify_model_att}
    \E[Y(0) \mid X = x, U = u, Z =1] = f(x) + \beta_{u0} \cdot u.
\end{equation}

Then, we define the bias to be the difference between the true expected value of control potential outcomes for treated units $\mu_{01}$ and the IPW estimand. We decompose the bias into (i) the strength of the unmeasured confounder $U$ in predicting $Y(0)$ for treated units beyond the observed covariates, $\beta_{u0}$ and (ii) the imbalance in $U$, $\delta_{u0}$:
\[
\E[Y(0)\mid Z = 1] - \E\left[\frac{1-Z}{\mathbb{P}(Z=1)}\frac{\pi(X)}{1-\pi(X)}Y\right] = \beta_{u0} \cdot \underbrace{\left(\E\left[U \mid Z = 1\right] - \E\left[\frac{1-Z}{\mathbb{P}(Z=1)}\frac{\pi(X)}{1-\pi(X)} U\right]\right)}_{\delta_{u0}}.
\]

Next, we derive upper and lower bounds for this product by using the partial identification of $\mu_{01}$ under the marginal sensitivity model:
\[
\inf_{h \in \calH(\Lambda^\ast)} \mu_{01}^{(h)} - \E\left[\frac{1-Z}{\mathbb{P}(Z=1)}\frac{\pi(X)}{1-\pi(X)}Y\right] \leq \beta_{u0} \cdot \delta_{u0} \leq \sup_{h \in \calH(\Lambda^\ast)} \mu_{01}^{(h)} - \E\left[\frac{1-Z}{\mathbb{P}(Z=1)}\frac{\pi(X)}{1-\pi(X)}Y\right].
\]
Finally, we construct finite-sample versions of these population bounds by bounding the bias as the maximum of the absolute values of the highest and lowest possible differences in the estimated values,
\[
|\beta_{u0} \cdot \delta_{u0}| \leq \max\left\{\left|\inf_{h \in \calH(\Lambda^\ast)} \hat{\mu}_{01}^{(h)} - \hat{\mu}_{01}\right|, \left|\sup_{h \in \calH(\Lambda^\ast)} \hat{\mu}_{01}^{(h)} - \hat{\mu}_{01}\right|\right\}.
\]


\end{document}